\documentclass[
superscriptaddress,
floatfix,
twocolumn,
]{revtex4-2}
\usepackage[utf8]{inputenc}
\usepackage{amsmath}
\usepackage{mathtools}
\usepackage{xcolor}
\usepackage{subcaption}
\usepackage{verbatim}
\usepackage{float}
\usepackage{placeins}
\usepackage[normalem]{ulem}
\usepackage{xcolor}

\newcommand{\csout}{\bgroup\markoverwith{\textcolor{red}{\rule[0.5ex]{2pt}{2pt}}}\ULon}

\begin{document}

\title{Predicting polarizabilities of silicon clusters using local chemical environments}

\author{Mario G. Zauchner}
\affiliation{Department of Materials, Thomas Young Centre, Imperial College London, South Kensington Campus, London, SW7 2AZ, United Kingdom}

\author{Stefano {Dal~Forno}}
\affiliation{Department of Physics, Imperial College London, South Kensington Campus, London, SW7 2AZ, United Kingdom}

\author{G\'abor C\'sanyi}
\affiliation{Engineering Laboratory, University of Cambridge, Trumpington Street, Cambridge CB2 1PZ, United Kingdom}
 
\author{Andrew Horsfield$^*$} 
\affiliation{Department of Materials, Thomas Young Centre, Imperial College London, South Kensington Campus, London, SW7 2AZ, United Kingdom}

\author{Johannes Lischner}
\affiliation{Department of Materials, Thomas Young Centre, Imperial College London, South Kensington Campus, London, SW7 2AZ, United Kingdom}
 
\date{August 2021}

\captionsetup[subfigure]{position=top, labelfont=bf,textfont=normalfont,singlelinecheck=off,justification=raggedright}

\begin{abstract}
Calculating polarizabilities of large clusters with first-principles techniques is challenging because of the unfavorable scaling of computational cost with cluster size. To address this challenge, we demonstrate that polarizabilities of large hydrogenated silicon clusters containing thousands of atoms can be efficiently calculated with machine learning methods. Specifically, we construct machine learning models based on the smooth overlap of atomic positions (SOAP) descriptor and train the models using a database of calculated random-phase approximation polarizabilities for clusters containing up to 110 silicon atoms. We first demonstrate the ability of the machine learning models to fit the data and then assess their ability to predict cluster polarizabilities using k-fold cross validation. Finally, we study the machine learning predictions for clusters that are too large for explicit first-principles calculations and find that they accurately describe the dependence of the polarizabilities on the ratio of hydrogen to silicon atoms and also predict a bulk limit that is in good agreement with previous studies.
\end{abstract}

\maketitle 

\section{Introduction}

Clusters and nanoparticles are used in a variety of scientific and industrial applications, including optoelectronics~\cite{Wang2003}, photocatalysis~\cite{Curtis2021}, medical imaging~\cite{Park2009, O'Farrell2006} or single electron transistors~\cite{FU2000177}. Electronic excitations often play a key role in these applications, but theoretical techniques for calculating excited-state properties of materials, such as the first-principles GW/Bethe-Salpeter method, are typically limited to very small systems. A key bottleneck of such excited-state calculations of clusters and nanoparticles is the determination of the static polarizability which is often calculated using a sum-over-states technique~\cite{Onida2002}.

As an efficient alternative to first-principles techniques, machine learning (ML) based techniques have been explored in recent years. For example, ML has been employed to efficiently represent potential energy surfaces~\cite{bartok_gaussian_2010, hansen_machine_2015, Kermode, faber_machine_2016} or to predict electronic ground state densities~\cite{brockherde_bypassing_2017, alred_machine_2018,Grifasi2019, chandrasekaran_solving_2019}. Additionally, projects such as the Materials Project \cite{Jain2013} and the Open Quantum Materials database\cite{Kirklin2015, Saal2013} have made an effort to make first-principles data of a wide range of materials publicly available. A key ingredient in ML methods is a descriptor which acts as a molecular fingerprint and encodes the structure and chemistry of a molecule. For example, the smooth-overlap of atomic positions (SOAP) descriptor~\cite{Bartok2013} has been widely used for the comparison of different chemical environments of an atom. For this, the overlap of the corresponding neighbourhood densities (constructed as a sum of Gaussians centered on atoms in the local environment) is expressed in terms of the coefficients in a basis of spherical harmonics and radial basis functions~\cite{Bartok2013}. Very recently, several groups have also started to explore the applicability of ML approaches to calculate molecular polarizabilities and dipole moments~\cite{Wilkins3401, Grisafi2018, Tuan-Anh2020, Ceriotti2020}. For example, Grisafi et. al. \cite{Grisafi2018} introduced a symmetry adapted variant of the SOAP descriptor~\cite{Bartok2013} to predict polarizability tensors of molecules. Similarly, Wilkins and coworkers~\cite{Wilkins3401} used the symmetry-adapted SOAP kernel to predict polarizabilities and first hyperpolarizabilities of small organic molecules with high accuracy. Recently, Ceriotti et. al. \cite{Ceriotti2020} used a combination of the symmetry adapted SOAP kernel and the scalar SOAP kernel to predict dipole moments of small molecules with close to DFT accuracy. However, the applicability of ML approaches to polarizabilities of clusters and nanoparticles , in particular the ability to make predictions about nanoparticles too large for first-principles calculations remains largely unexplored.  \\ 

To assess the performance of ML approaches for cluster polarizabilities, we focus in this work on hydrogenated silicon clusters. These systems are well suited for this purpose because their polarizabilities have been studied in detail with a variety of modelling techniques. For example, simple empirical models, such as bond polarizability models, have been used to predict Raman spectra of silicon clusters in good agreement with experiment~\cite{Povarnitsyn2020}. Empirical models can be extended beyond the assumption of additivity of atomic polarizabilities. A class of models that captures interactions between polarization centres are dipole interaction models \cite{Wang2011}, which have been successfully applied to the construction of polarizable force fields \cite{Wang2006}. Highly accurate cluster polarizabilities can be obtained using ab initio approaches such as density functional theory (DFT)~\cite{Vasilev1997, Deng2000,Bazterra2002, Jackson1999, Jackson2005,Maroulis2003}, M\o{}ller-Plesset perturbation theory\cite{Maroulis2003}, coupled-cluster theory ~\cite{Maroulis20037, Papadopoulos2006, Maroulis2003} or the random phase approximation (RPA)~\cite{fetter2003quantum, Jansik, MOCHIZUKI2001451}. For example, Mochizuki and Agren~\cite{MOCHIZUKI2001451} used the RPA and the second-order polarization propagator approximation to calculate the polarizabilities of spherical hydrogenated silicon clusters with up to 35 Si atoms and found that the polarizability per silicon atom approaches the bulk limit from below. In contrast, for unhydrogenated silicon clusters Jackson and coworkers found that the bulk value is approached from above as the size of the cluster increases \cite{Jackson1999, Jackson2005}. This behaviour was attributed to the presence of dangling bonds on the surface. Furthermore, it was observed that the polarizability depends sensitively on the shape of the cluster~\cite{Jansik, Jackson2005}. Jansik et al.~\cite{Jansik} compared polarizabilities of three-dimensional (3D), two-dimensional (2D) and one-dimensional (1D) hydrogenated silicon structures and found that the presence of $\pi$-bonds in 2D systems leads to a much stronger increase in the polarizability as a function of cluster size when compared to 1D and 3D clusters \cite{Jansik}. A similar trend was observed when comparing prolate and compact clusters, with prolate structures showing a significantly larger polarizability per silicon atom than compact structures \cite{Jackson2005}.

In the present work, we explore the ability of machine learning models based on the Smooth Overlap of Atomic Positions (SOAP)~\cite{Bartok2013} descriptor to describe and predict static polarizabilities of hydrogenated silicon clusters calculated from random phase approximation (RPA) static density-density response functions. We chose the SOAP descriptor due it its widespread use for a variety of atomic scale regression tasks and its systematic nature. Previous work \cite{Wilkins3401, Grisafi2018} has already demonstrated the ability to predict isotropic scalar polarizabilities and also the full polarizability tensor using SOAP and generalizations thereof. The symmetry-adapted SOAP descriptor has also been used successfully in conjunction with physical insights~\cite{Ceriotti2020} to predict molecular dipole moments. Furthermore, we note that SOAP is a generic 3-body descriptor of the neighbour density that obeys rotational and permutational symmetries \cite{Musil2021}, and as such encompasses simpler descriptors such as RDFs and ADFs \cite{Honrao2020, Choudhary2018, rohrhofer2021importance} (which are particular projections of the neighbour density), and in the limit of no basis truncation equivalent to other 3-body descriptors such as Behler-Parrinello Atom Centered Symmetry Functions \cite{Behler} and the FCHL \cite{Faber2018} descriptors.. To generate a data set, we first calculate scalar isotropic polarizabilities of a set of clusters containing between 10 and 110 silicon atoms using the RPA. We then investigate the ability of the ML approach to reproduce the calculated polarizabilities and find that almost perfect agreement can be obtained when the size of the local chemical environments is sufficiently large to contain the whole cluster. Importantly, the ML models already describe the qualitative behaviour of the average polarizability per atom if the local environment only contains nearest neighbour atoms. These findings establish the fittability of RPA scalar polarizabilities using local SOAP descriptors which - in contrast to mean-field DFT data - has not been explored to date. Next, we study the ability of ML to predict polarizabilities of clusters. Interestingly, we find that the predictive power of ML is strongest when the size of the chemical environment is relatively small. These insights enable the reliable prediction of polarizabilities of large clusters which are difficult to calculate with standard first-principles techniques and constitute a first step towards efficient ML approaches for excited-state properties of materials.

\section{Methods}

\subsection{Random Phase Approximation polarizabilities}
Scalar polarizabilities of molecules and clusters were calculated within the RPA in a linear response framework. The RPA was chosen because it is known to give an accurate description of the dielectric properties of bulk silicon~\cite{Hybertsen1987}. The polarizability tensor $\alpha_{ij}$ relates the induced dipole moment with Cartesian components $\mu_i$ to the applied static electric field $E_j$ according to   

\begin{equation}
\mu_i = \sum_j\alpha_{ij}E_j. 
\end{equation}

To obtain an expression for $\alpha_{ij}$, we express $\mu_i$ in terms of the induced electronic charge density $\Delta \rho(\mathbf{r})$ via

\begin{equation}
\mu_i = -e\int d\mathbf{r}\Delta\rho(\mathbf{r})r_i,
    \label{dipolemoment}
\end{equation}
where $e$ denotes the proton charge and $r_i$ is the Cartesian component of the position vector. The induced charge density is determined by the interacting density-density response function $\chi(\mathbf{r},\mathbf{r}')$ according to 

\begin{equation}
    \Delta \rho(\mathbf{r}) = e \sum_j E_j \int d\mathbf{r}' \chi(\mathbf{r},\mathbf{r}') r'_j,
\end{equation}
where we used that the potential associated with the applied electric field is given by $V(\mathbf{r})=e\sum_j E_j r_j$. Combining these equations yields
\begin{equation}
\alpha_{ij} = -e^2 \int d\mathbf{r}d\mathbf{r'} \chi(\mathbf{r,r'})r_i r'_j.
    \label{eq:alphaij}
\end{equation}
Finally, the scalar polarizability $\alpha$ is obtained by dividing the trace of $\alpha_{ij}$ by three.

To evaluate Eq.~\eqref{eq:alphaij} the interacting density-density response function must be determined. In the RPA $\chi$ obeys the Dyson equation
\begin{align}
        \chi(\mathbf{r},\mathbf{r}')=&\chi_0(\mathbf{r},\mathbf{r}') + \\ &\int d\mathbf{r}_1 d\mathbf{r}_2 \chi_0(\mathbf{r},\mathbf{r}_1) v(\mathbf{r}_1-\mathbf{r}_2) \chi(\mathbf{r}_2,\mathbf{r}'),
\end{align}
where $v(\mathbf{r}-\mathbf{r}')$ denotes the Coulomb interaction and $\chi_0$ is the non-interacting density-density response function given by~\cite{Adler,Wiser} 
\begin{align}
        \chi_0(\mathbf{r},\mathbf{r}')=&\sum_{ij} \frac{f_i(1-f_j)}{\epsilon_i-\epsilon_j} \times \\ &\left[ \phi_i^*(\mathbf{r})\phi_j(\mathbf{r})  \phi^*_j(\mathbf{r}')\phi_i(\mathbf{r}') + \text{c.c.} \right],
        \label{adlerwiser}
\end{align}
where $f_i$ denotes an occupancy factor and $\phi_i$ and $\epsilon_i$ denote Kohn-Sham orbitals and eigenvalues, respectively. Note that the summation ranges over both occupied and unoccupied states resulting in the well-known difficulties of converging such sum-over-states expressions.
To numerically calculate scalar polarizabilities, we employ a plane-wave/pseudopotential approach. Specifically, the BerkeleyGW programme package~\cite{bgw1,bgw2} is used to calculate $\chi_{\mathbf{G}\mathbf{G}'}$ where $\mathbf{G}$ and $\mathbf{G}'$ denote reciprocal lattice vectors of the periodically repeated supercell. Note that interactions between images are avoided by using a truncated Coulomb interaction. The interacting density-density response function in real space is then given by 
\begin{equation}
    \chi(\mathbf{r,r'}) = \frac{1}{V} \sum_{\mathbf{G,G'}}e^{i\mathbf{G\cdot r}} \chi_{\mathbf{G,G'}}e^{-i\mathbf{G'\cdot r'}},
    \label{rchi}
\end{equation}
where $V=L^3$ denotes the volume of the cubic supercell, with $L$ being the side length. Finally, the scalar polarizability is found to be
\begin{equation}
    \alpha = \frac{e^2}{3V} \sum_i \sum_{\mathbf{G,G'}} \chi_{\mathbf{G,G'}} \Delta_{\mathbf{G},i} \Delta_{\mathbf{G'},i}^*, 
    \label{alpha}
\end{equation}
with 
\begin{equation}
\Delta_{\mathbf{G},x} =
\begin{cases}
   \frac{L^4}{2}\delta_{G_x,0}\delta_{G_y,0}\delta_{G_z,0} & \text{if } G_x=0,\\
     \frac{L^3}{ i G_x}\delta_{G_y,0}\delta_{G_z,0}              & \text{otherwise},
\end{cases}
\end{equation}
and similar expressions for $\Delta_{\mathbf{G},y}$ and $\Delta_{\mathbf{G},z}$.

Finally, we note that other - more efficient - approaches than the one described above exist for the calculation of the static scalar polarizability, such as the finite field method~\cite{Calaminici1998}. However, our ultimate interest is in applying ML techniques to accelerate excited-state calculations and these methods require the full interacting density-density response function which cannot easily be obtained with other methods.

\subsection{Environment descriptors}

The ability to assess the similarity of different chemical environments plays a key role in machine learning of material properties. In this work, we use the SOAP approach~\cite{Bartok2013} where the environment of atom $i$ is described by the set of neighbourhood densities
\begin{equation}
    \rho_{i}^{\nu}(\mathbf{r})=\sum_{i=j}^{N_\nu} e^{-\gamma_\nu ( \mathbf{r}-\mathbf{r}_{ij} )^2}, 
    \label{density}
\end{equation}
where $\nu$ denotes a specific element that is present in the atom's environment with $N_\nu$ being the number of such atoms up to a given cut-off radius $r_{cut}$. In addition, $\gamma_\nu$ is a hyperparameter describing the size of the neighbour atom. 

The similarity of two chemical environments described by the neighbourhood densities $\rho_{i}=\{\rho_{i}^\nu\}_{\nu}$ and $\rho_{j}=\{{\rho_j}^\nu\}_\nu$ can be measured by the kernel~\cite{Ceriotti2018}
\begin{equation}
    k(\rho_i,\rho_j)= \int d\hat{R} \bigg \vert \sum_{\nu}  \int d\mathbf{r} \rho_i^\nu(\mathbf{r}) \rho_j^{\nu}(\hat{R}\mathbf{r}) \bigg \vert^2,
    \label{soapkernel}
\end{equation}
where $\hat{R}$ denotes a rotation matrix. To evaluate the kernel integral, the angular dependence of the neighbourhood densities is expanded in a basis of spherical harmonics $Y_{lm}$ and the radial part in a set of orthogonal radial basis functions $g_n(r)$ according to
\begin{equation}
    \rho_i^\nu(\mathbf{r})= \sum_{nlm} c^{\nu}_{i,nlm}g_n(r) Y_{lm}(\mathbf{\hat{r}}),
\end{equation}
where $c^{\nu}_{i,nlm}$ is an expansion coefficient. Here, $l$ ranges from zero to a cut-off value $l_\text{max}$ and $m$ ranges from $-l$ to $l$. As radial basis functions, the modified spherical Bessel functions of the first kind are used and $n$ ranges from zero to a cut-off value $n_\text{max}$.

The similarity kernel Eq.~\eqref{soapkernel} has the appealing property that the integrals can be carried out analytically yielding~\cite{Bartok2013}
\begin{equation}
    \begin{multlined}
    k(\rho_i,\rho_j)= \sum_{\nu \leq \nu'} \sum_{nn'l} d^{\nu,\nu'}_{i,nn'l}d^{\nu,\nu'}_{j,nn'l}
    \label{soapkernel2}
    \end{multlined}
\end{equation}

\begin{equation}
\begin{multlined}
d^{\nu,\nu'}_{i,nn'l}= \sum_{m} c^{\nu}_{i,nlm} (c^{\nu'}_{i,n'lm})^*.
\end{multlined}
\label{descriptorvec}
\end{equation}
From the above expressions, it can be seen that the set of coefficients $\{d^{\nu,\nu'}_{i,nn'l} \}$ plays the role of a descriptor vector $\mathbf{d}_i$ for the environment of atom $i$. In practice, we calculate the descriptor vectors using the Quippy software package~\cite{jkermode}. The kernel matrix $ k(\rho_i,\rho_j)$ is then calculated according to 

\begin{equation}
     k(\rho_i,\rho_j) = \mathbf{d}_i \cdot \mathbf{d}_j.
     \label{kernel}
\end{equation}

Finally, we note that the sensitivity of the kernel to differences between atomic environments can be increased by defining the effective SOAP kernel~\cite{Bartok2013,Ceriotti2018}
\begin{equation}
    K(\rho_i,\rho_j)=\bigg ( \frac{k(\rho_i,\rho_j)}{\sqrt{k(\rho_i,\rho_i)k(\rho_j,\rho_j)}} \bigg )^\epsilon.
    \label{effsoapkernel}
\end{equation}
In this work, we use $\epsilon=2$.

\subsection{Learning cluster polarizabilities}

The SOAP descriptor allows the comparison of different environments of a given atom. However, the polarizability is calculated for an entire molecule or cluster consisting of many atoms. To harness the SOAP approach for the prediction of cluster polarizabilities, it is therefore necessary to relate atomic properties to cluster properties. One way to achieve this is by expressing the polarizability $\alpha_I$ of cluster $I$ as the sum of atomic contributions $\alpha_i$ \cite{Ceriotti2018} according to

\begin{equation}
    \alpha_I = \sum_{i=1}^{N_I}\alpha_i,
    \label{eq:pol}
\end{equation}
where $N_I$ denotes the total number of atoms in the cluster. While the atomic contributions can provide some valuable intuition about the dielectric response of complex clusters, it is important to stress that these quantities are not directly measurable and should be interpreted with care \cite{Bart2017}.

Using standard kernel ridge regression, the atomic polarizabilities can be expressed as
\begin{equation}
    \alpha_i = \sum_j^{N_\text{train}} K_{ij} \zeta_j,
    \label{eq:alphai}
\end{equation}
where $N_\text{train}$ denotes the total number of atoms in the training set (i.e. the total number of atoms contained in all training set clusters), $\zeta_j$ is a coefficient obtained from training the SOAP model, and $K_{ij} \equiv K(\rho_i,\rho_j)$. Inserting Eq.~\eqref{eq:alphai} into Eq.~\eqref{eq:pol} yields
\begin{equation}
     \alpha_I = \sum_{i}^{N_I} \sum_j^{N_\text{train}} K_{ij} \zeta_j = \sum_j^{N_\text{train}} K^\text{sum}_{I,j} \zeta_j,
    \label{sumkernel}
\end{equation}
where we defined the sum kernel $K^\text{sum}_{I,j} = \sum_{i}^{N_I}  K_{ij}$.

Determining the coefficients $\zeta_j$ is difficult as the fit to the calculated cluster polarizabilities is strongly underdetermined (as the number of coefficients is the total number of atoms of all clusters in the training set). To make progress, the number of coefficients must be reduced. Intuitively, this should be possible as the atomic environments of many atoms in the training set are very similar. Practically, this sparsification is achieved by means of a singular value decomposition (SVD) of the descriptor matrix $\mathbf{D}$ whose rows contain the descriptor vectors from Eq.~\eqref{descriptorvec}. Specifically, $\mathbf{D}$ is expressed as
\begin{equation}
    \mathbf{D} = \mathbf{U} \mathbf{\Sigma} \mathbf{V}^T,
\end{equation}
where $\mathbf{U}$ and $\mathbf{V}^T$ contain the right and left singular vectors, respectively, and $\mathbf{\Sigma}$ is a diagonal matrix containing the singular values. If many environments in $\mathbf{D}$ are similar, only a few singular values will have large magnitudes. We only retain those singular values which are larger than a given threshold and use the corresponding left singular vectors (which form a matrix $\tilde{\mathbf{V}}$) as a new basis to represent $\mathbf{D}$. 

The elements of the SOAP kernel $\tilde{\mathbf{K}}$ corresponding to this new set of effective descriptors are obtained by projecting the descriptors $\mathbf{d}_i$ onto the rows $\tilde{\mathbf{v}}_j$ of the truncated matrix of singular vectors $\tilde{\mathbf{V}}$ according to 

\begin{equation}
   \tilde{K}_{ij} = \mathbf{d}_i \cdot \tilde{\mathbf{v}}_j.
    \label{sparse}
\end{equation}

Next, the effective sum kernel $\tilde{\mathbf{K}}^\text{sum}$ can be calculated using Eq.~\ref{sumkernel}, but now the number of coefficients $\zeta_i$ is equal to the number of singular vectors whose singular values exceed the threshold. Finally, the vector of coefficients $\zeta$ is obtained from~\cite{Ceriotti2018}

\begin{equation}
    \zeta = \left[ \mathbf{\tilde{V}}^T \mathbf{\tilde{V}} + (\tilde{\mathbf{K}}^{\text{sum}})^T \mathbf{\Lambda}^{-1} \tilde{\mathbf{K}}^{\text{sum}} \right]^{-1} (\tilde{\mathbf{K}}^{\text{sum}})^T \mathbf{\Lambda}^{-1}\boldsymbol\alpha,
    \label{zeta}
\end{equation}
where $\mathbf{\Lambda}= \lambda \mathbf{I}$ with $\lambda$ being a regularization parameter and $\boldsymbol\alpha$ denotes the vector of calculated cluster polarizabilities.

Alternatively, the cluster polarizability can be expressed as the number of silicon atoms multiplied by their average polarizability $\alpha^\text{av}$ (note that in this definition $\alpha^\text{av}$ also contains the smaller contribution from the hydrogen atoms)
\begin{equation}
    \alpha =  N_{Si} \alpha^\text{av}_{Si}.
\end{equation}

To calculate the average polarizability, we average the SOAP kernel matrix over environments belonging to pairs of clusters ~\cite{Bart2017}
\begin{equation}
    K^{\text{av}}_{IJ} = \frac{1}{N_I N_J}\sum_i^{N_I} \sum_j^{N_J} K_{ij}.
    \label{eq:Kav}
\end{equation}

Using kernel ridge regression, the average polarizability of the silicon atoms in a given cluster is expressed as 
\begin{equation}
    \alpha^\text{av}_{Si} = \sum_J^{n_\text{train}} K^\text{av}_{J} \zeta_J,
\end{equation}
where $n_\text{train}$ denotes the number of clusters in the training set and the vector of coefficients $\zeta_J$ is determined by
\begin{equation}
    \zeta = ( \mathbf{K}^\text{av} + \mathbf{\Lambda} )^{-1}\ \boldsymbol\alpha^{\text{av}},
    \label{squarekernel}
\end{equation}
where $\boldsymbol\alpha^\text{av}$ is the vector containing the average polarizabilities per atom of the training set clusters. As a consequence of the averaging, no additional sparsification procedure is required as in the case of the sum kernel.

This method has the advantage that the average polarizability can be written as a sum of atomic contributions, which allows one to assign polarizabilities to individual atoms. This can be achieved by omitting the average over the index $i$ in Eq.~\ref{eq:Kav}, which yields a prediction for each silicon atom in a cluster

It is interesting to note that the polarizability obtained from the average kernel can also be expressed as a sum of atomic contributions given by
\begin{equation}
    \alpha_i = \frac{1}{N_J} \sum_J^{n_\text{train}}\sum_j^{N_J} K_{ij}\zeta_J.
\label{eq:atomic}
\end{equation}

Apart from the scaling factor $1/N_J$, the last equation is very similar to Eq.~\eqref{eq:alphai} of the sum kernel approach, with the additional constraint that the coefficients $\zeta_j$ on atoms in a cluster $J$ are all equal, $\zeta_j = \zeta_J~\forall j\in J$. The effect of the scaling factor is that while the sum kernel is extensive (its magnitude scales with the number of atoms in the cluster), the average kernel is intensive, independent of system size. As a consequence of this, large clusters get more heavily weighted in the solution of the least squares problem, Eq.~\eqref{zeta}, compared with that for the average kernel. 

Finally, we also use the ``coherent average'' kernel (denoted ``coh''), which is obtained as follows. Rather than computing a SOAP descriptor for each atomic environment, as in Eq.~\eqref{soapkernel2}, we take the spherical harmonic coefficients $c^{\nu}_{nlm}$ and average them first to obtain, for cluster $I$ 
\begin{equation}
    {\bar c}^{\nu}_{I,nlm} = \frac{1}{N_I} \sum_{i=1}^{N_I} c^{\nu}_{i,nlm},
\end{equation}
and then square these to form the averaged descriptor vector $\mathbf{\bar d}_I$ with components,
\begin{equation}
    {\bar d}^{\nu,\nu'}_{I,nn'l} = \sum_{m} {\bar c}^\nu_{I,nlm} ({\bar c}^{\nu'}_{I,n'lm})^*,
\end{equation}
and the coherent (unnormalized) kernel between clusters $I$ and $J$ as
\begin{equation}
    k^{\rm coh}(I,J) = \mathbf{{\bar d}}_I\cdot \mathbf{{\bar d}}_{J}.
\end{equation}

\subsection{Physical models}
We also use two simple physical-based models to fit the calculated RPA polarizabilities. In the first approach, the cluster polarizability is assumed to be proportional to the number of silicon atoms $N_{Si}$ in the cluster, i.e. 
\begin{equation}
    \alpha = \alpha^\text{av}_{Si} N_{Si},
    \label{linear}
\end{equation}
with $\alpha^{av}_{Si}$ denoting the average polarizability per silicon atom (again, any contributions from hydrogen atoms is included in $\alpha^{av}_{Si}$ in this definition). In contrast to the SOAP fitting with the average kernel, the average polarizability is assumed to be the same for all clusters. 

The second model is a bond polarizability approach where the cluster polarizability is expressed as a sum of contributions from Si-Si bonds and Si-H bonds according to
\begin{equation}
    \alpha =  \alpha_{Si-Si} N_{Si-Si}+  \alpha_{Si-H} N_{Si-H},
    \label{bondorder}
\end{equation}
where $\alpha_{Si-H}$ and $\alpha_{Si-Si}$ are the polarizabilities of Si-H and Si-Si bonds, respectively, and $N_{Si-Si}$ and $N_{Si-H}$ are the number of Si-Si and Si-H hydrogen bonds, respectively. While this model explicitly includes the contribution of the hydrogen atoms, it is also assumed that the bond polarizabilities are independent of the cluster size and shape. 

In a hydrogenated silicon cluster with only sp$^3$ bonding, the number of Si-Si bonds and Si-H bonds can be expressed in terms of the number of silicon and hydrogen atoms. In particular, $N_{Si-Si}$ is given by $(4N_{Si}-N_H)/2$, and $N_{Si-H}$ is equal to $N_H$. Substituting these expressions into Eq.~\ref{bondorder} yields
\begin{equation}
    \alpha = \frac{4N_{Si}-N_H}{2} \alpha_{Si-Si} + N_H\alpha_{Si-H}.
\end{equation}
Dividing both sides by $N_{Si}$ yields the polarizability per silicon atom 
\begin{multline}
    \frac{\alpha}{N_{Si}}= 2\alpha_{Si-Si} + \left( \alpha_{Si-H}-\frac{\alpha_{Si-Si}}{2} \right) \frac{N_H}{N_{Si}}.
    \label{eq:alphaNSi}
\end{multline}
Interestingly, this shows that the polarizability per silicon atom is a function of the ratio of hydrogen and silicon atoms only.

\subsection{Generation of clusters}
To generate atomic structures of hydrogenated silicon clusters we follow a similar procedure as Barnard et al.~\cite{barnard2015silicon} who carve spherical clusters from a perfect silicon crystal, terminate the dangling bonds on the surface with hydrogen atoms and then relax the atomic positions using DFT. Unfortunately, this approach only yields very few clusters with 100 or less silicon atoms. Because of the relatively large computational cost associated with the RPA polarizability calculations, we instead use the following approach to generate clusters: starting from the spherical Si$_{123}$H$_{100}$ cluster, we remove silicon atoms from the surface, terminating any dangling bonds with hydrogen atoms and relax the structure with DFT. In this way, a set of 100 hydrogenated silicon clusters containing between 10 and 110 Si atoms is obtained for which RPA polarizabilities are calculated. In addition, we include the spherical clusters with less than 123 Si atoms from Barnard et al.~\cite{barnard2015silicon}.

\subsection{Computational details}
The plane-wave/pseudopotential DFT code Quantum Espresso \cite{Giannozzi_2009, Giannozzi_2017} was used to obtain Kohn-Sham energies $\epsilon_n$ and wavefunctions $\phi_n(\mathbf{r})$. We employed the PBE exchange-correlation functional, norm-conserving pseudopotentials from the original Quantum Espresso Pseudopotential library \cite{Giannozzi_2009, Giannozzi_2017} and a plane-wave cut-off of 65 Ry. The clusters were placed in a cubic unit cell with sufficient vacuum to avoid interactions between periodically repeated images. Next, cluster polarizabilities were calculated with BerkeleyGW~\cite{bgw1,bgw2} using a plane-wave cutoff of 6 Ry and a truncated Coulomb interaction. A total of 600 Kohn-Sham states were included in the summation for $\chi$ which was found to be sufficient to converge the scalar polarizabilities. 
SOAP descriptors were constructed with $l_{max}=9$ and $n_{max}=20$ and $\gamma_{\nu}=2.0$ for $r_{cut}\le 10.0 \text{\AA}$ and $\gamma_{\nu} = 0.5$ for $r_{cut}> 10.0$\AA. In all calculations, we only study local environments of silicon atoms. As all hydrogen atoms are bonded to silicon atoms, their contribution to the cluster polarizabilities can be captured indirectly through their influence on the silicon atoms.

\section{Results and Discussion}

\subsection{Fitting polarizabilities}

\begin{figure}[htbp!]
    \centering
     \begin{subfigure}[b]{0.45\textwidth}
     \caption{}
    \includegraphics[scale=0.48]{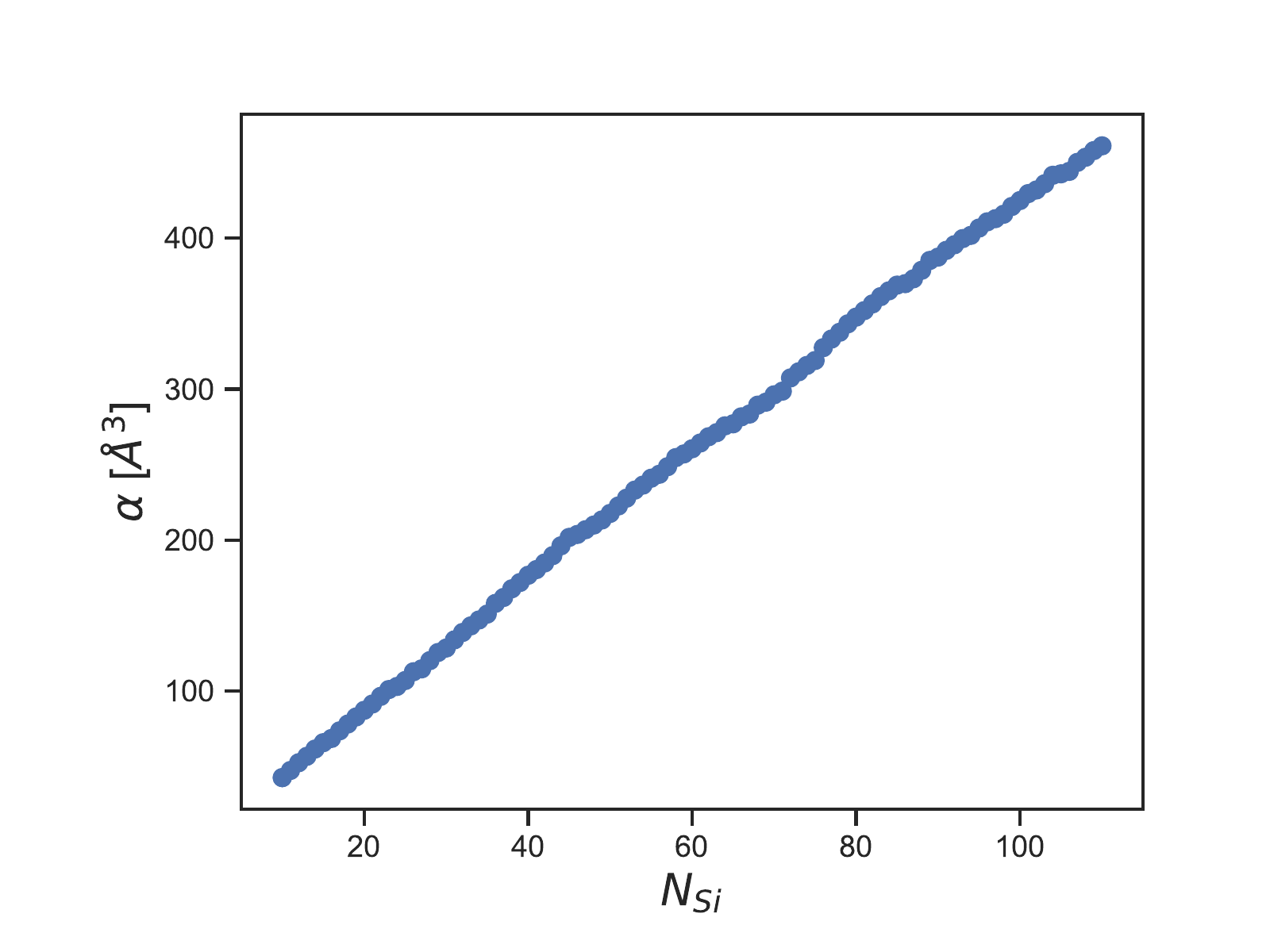}
    \label{pol}
    \end{subfigure}
    \begin{subfigure}[b]{0.45\textwidth}
     \caption{}
    \includegraphics[scale=0.48]{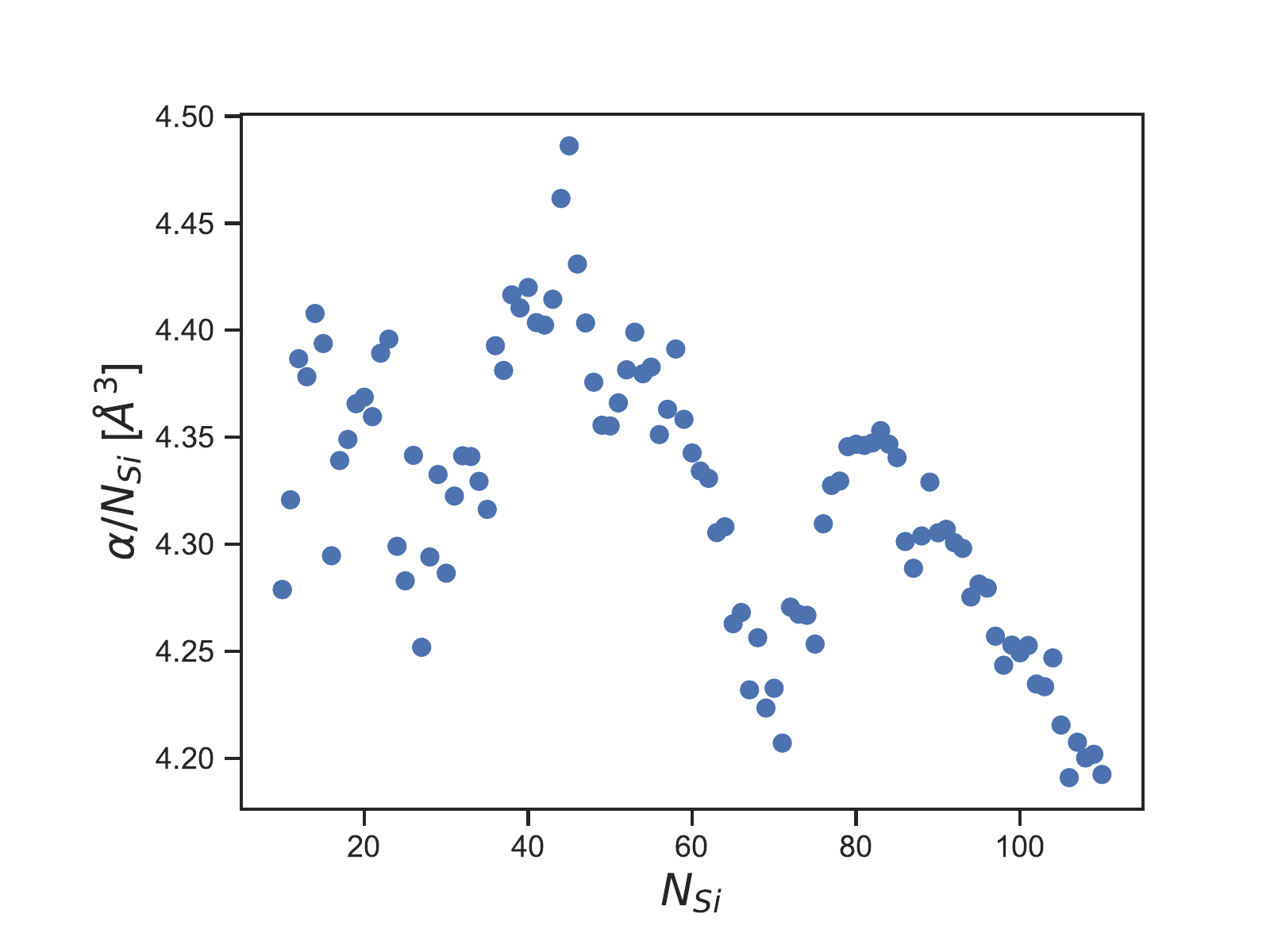}
    \label{avpoln}
    \end{subfigure}
    \caption{(a) RPA scalar polarizability $\alpha$ of hydrogenated silicon clusters versus number of silicon atoms $N_{Si}$. The polarizability increases approximately linearly with $N_{Si}$. (b) RPA polarizability divided by $N_{Si}$ shows deviations from linearity.}
    \label{polvnsi}
\end{figure}

Fig.~\ref{polvnsi}(a) shows the RPA polarizabilities of the hydrogenated silicon clusters as function of the number of silicon atoms in the cluster. We observe that the polarizability exhibits a linear behaviour which suggests that the Si atoms provide the dominant contribution.

Deviations from the linear behaviour become explicit when the cluster polarizability is divided by the number of silicon atoms, see Fig.~\ref{polvnsi}(b). For clusters containing more than 80 Si atoms, the polarizability per silicon atom decreases. Interestingly, $\alpha/N_{Si}$ increases for cluster between 70 and 80 silicon atoms, but decreases again for clusters between 40 and 70 silicon atoms. For clusters with less than 40 Si atoms, there is a significant amount of scatter in the polarizabilities but overall $\alpha/N_{Si}$ tends to increase with increasing number of Si atoms. Overall, the polarizability per silicon atom has an M-like shape as function of the number of silicon atoms. 

For very large clusters, $\alpha/N_{Si}$ should converge to the atomic RPA polarizability of bulk silicon which is 3.77~\AA$^3$ (determined using the Clausius-Mossotti relation using a bulk dielectric constant of $12.2$~\cite{Hybertsen1987}). This explains the observed decrease of $\alpha/N_{Si}$ for $N_{Si}>80$. Note that in our results the bulk value is not approached from below because we have not removed the hydrogen contributions from the cluster polarizabilities~\cite{MOCHIZUKI2001451,Jansik}.

\begin{figure}[htbp!]
    \centering
    \captionsetup[subfigure]{labelformat=empty}
    \begin{subfigure}[b]{0.0\textwidth}
   \caption{\label{avpol}}
    \end{subfigure}
    \captionsetup[subfigure]{labelformat=empty}
    \begin{subfigure}[b]{0.0\textwidth}
    \caption{\label{avpolSOAP}}
    \end{subfigure}
    \captionsetup[subfigure]{labelformat=empty}
     \begin{subfigure}[b]{0.48\textwidth}
    \includegraphics[width=0.99\textwidth]{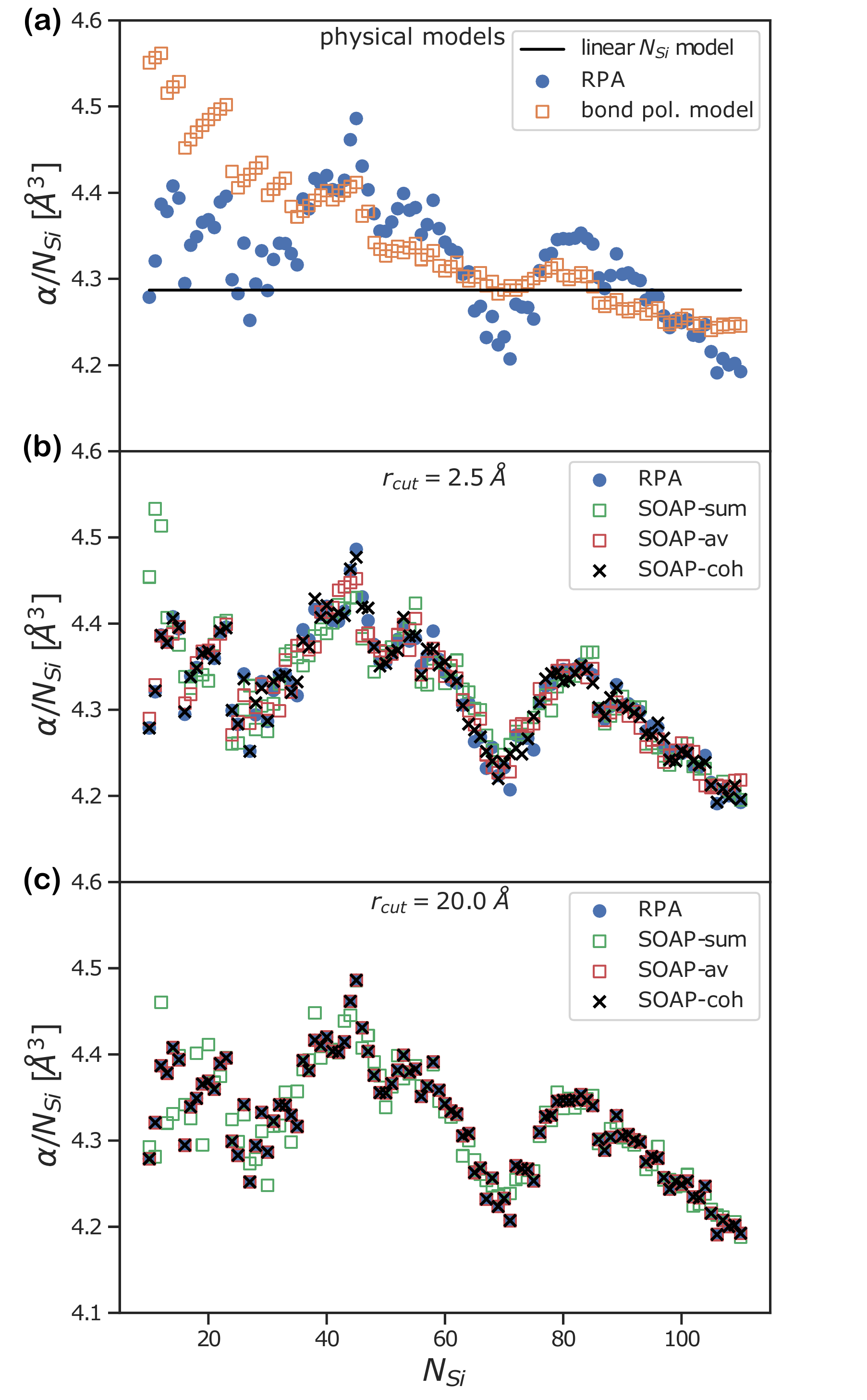}
    \caption{\label{avpolSOAPlarge}}
    \end{subfigure}
    \caption{Comparison of $\alpha/N_{Si}$ (cluster polarizability divided by the number of silicon atoms) from RPA calculations (blue dots) with fits from various models. (a) The black line shows the optimal fit for the linear $N_{Si}$ model, see Eq.~\ref{linear}. Orange triangles denote results from the bond polarizability model, see Eq.~\ref{bondorder}. (b) Results obtained using SOAP with the sum kernel (green squares), the average kernel (red squares) and the coherent kernel (black crosses). A cut-off of $2.5$ \AA\; was employed. (c) Same as (b) but with a cut-off of $20$ \AA.}
    \label{polfit}
\end{figure}

To understand these findings, we first compare our results to two physical-based models: a model in which the cluster polarizability is assumed to be proportional to the number of Si atoms (denoted the linear $N_{Si}$ model) and a bond polarizability model (see Methods). The parameters of both models were fitted to the calculated RPA data using a least squares optimization. The results are shown in Fig.~\ref{polfit} (a). While the linear $N_{Si}$ model cannot capture any dependence of $\alpha/N_{Si}$ on the number of silicon atoms, the bond polarizability model correctly describes several key features. In particular, it shows a decreasing trend for large clusters and a minimum near $N_{Si}=70$. For small clusters, the bond polarizability model predicts an increase in polarizability as the number of Si atoms is reduced in disagreement with the RPA data. Interestingly, the bond polarizability model also features a significant scatter for small clusters. As discussed in the methods section, $\alpha/N_{Si}$ in the bond polarizability model only depends on the ratio of hydrogen and silicon atoms $N_H/N_{Si}$ suggesting that this parameter is an important effective descriptor of the hydrogenated silicon clusters.

While the bond polarizability model captures several features, we note that neither of the two physical models can capture the full M-shape of the polarizability per Si atom in Fig.~\ref{polfit} (a).
Furthermore, from the least square fits of the linear $N_{Si}$ model to the RPA data, we find $\alpha^\text{av}_{Si}=4.29$~$\text{\AA}^3$. This is significantly larger than the RPA value in bulk Si of 3.77~$\text{\AA}^3$. The parameters of the bond polarizability model are found to be $\alpha_{Si-Si}=1.98$~$\text{\AA}^3$ and $\alpha_{Si-H}=1.32$~$\text{\AA}^3$. As the polarizability per Si atom is $2\alpha_{Si-Si}$, the predicted bulk value is 3.96~$\text{\AA}^3$ which is in better agreement with RPA results.

The above analysis demonstrates that both physical-based models have several shortcomings. This is a consequence of two factors: (i) their parameters do not depend on the properties of the local chemical environment, i.e. bond lengths or bond angles. 
In particular for small clusters, significant atomic relaxations occur resulting in changes to the bond polarizabilites compared to the larger clusters which are not captured by the bond polarizability model.
(ii) The models do not capture the effects of interactions between the polarizable units. As a consequence, they cannot distinguish between clusters containing the same numbers of Si and H atoms and do not capture the dependence of the polarizability on the cluster shape. To overcome these problems, we now explore the ability of machine learning models to describe the polarizabilities of Si clusters.

Figures~\ref{polfit} (b) and \ref{polfit} (c) show the results from the machine learning model using both the sum kernel, the average kernel and the coherent kernel (see methods). The real space cutoff that determines the size of the chemical environment of each atom is $r_c=2.5$~\AA\; in Fig.~\ref{polfit}(b) and $r_c=20$~\AA\; in Fig.~\ref{polfit}(c). In the fit, the regularization parameter $\lambda$ was kept small ($10^{-15}$ for the sum kernel model and $10^{-12}$ for the average and coherent kernels) in order to allow as much flexibility in the parameters as possible. For the smaller cut-off (where only nearest neighbour atoms are included in the local environment), all three kernels provide an improved description compared to the physical-based models. Specifically, they capture the M-shape of $\alpha/N_{Si}$ as function of $N_{Si}$ and also reproduce the scatter for smaller clusters. 
The coherent kernel is slightly better than the averaged kernel, and significant deviations from the calculated polarizabilities are only observed for the smallest cluster sizes when the sum kernel is used. When $r_c$ is increased to 20~\AA, the agreement between the ML models and the calculated polarizabilities is significantly improved. In particular, the results from the average and the coherent kernel are in almost perfect agreement with the data, while the sum kernel results show small deviations for smaller clusters. The good results obtained for the short cutoff indicate that polarizabilities are dominated by local chemical effects. However, long-range interactions also influence polarizabilities and this is captured when the cutoff radius is increased.

\subsection{Predicting polarizabilities}

\begin{figure}
    \centering
    \begin{subfigure}{0.48\textwidth}
       \caption{}
    \includegraphics[width=1.0\textwidth]{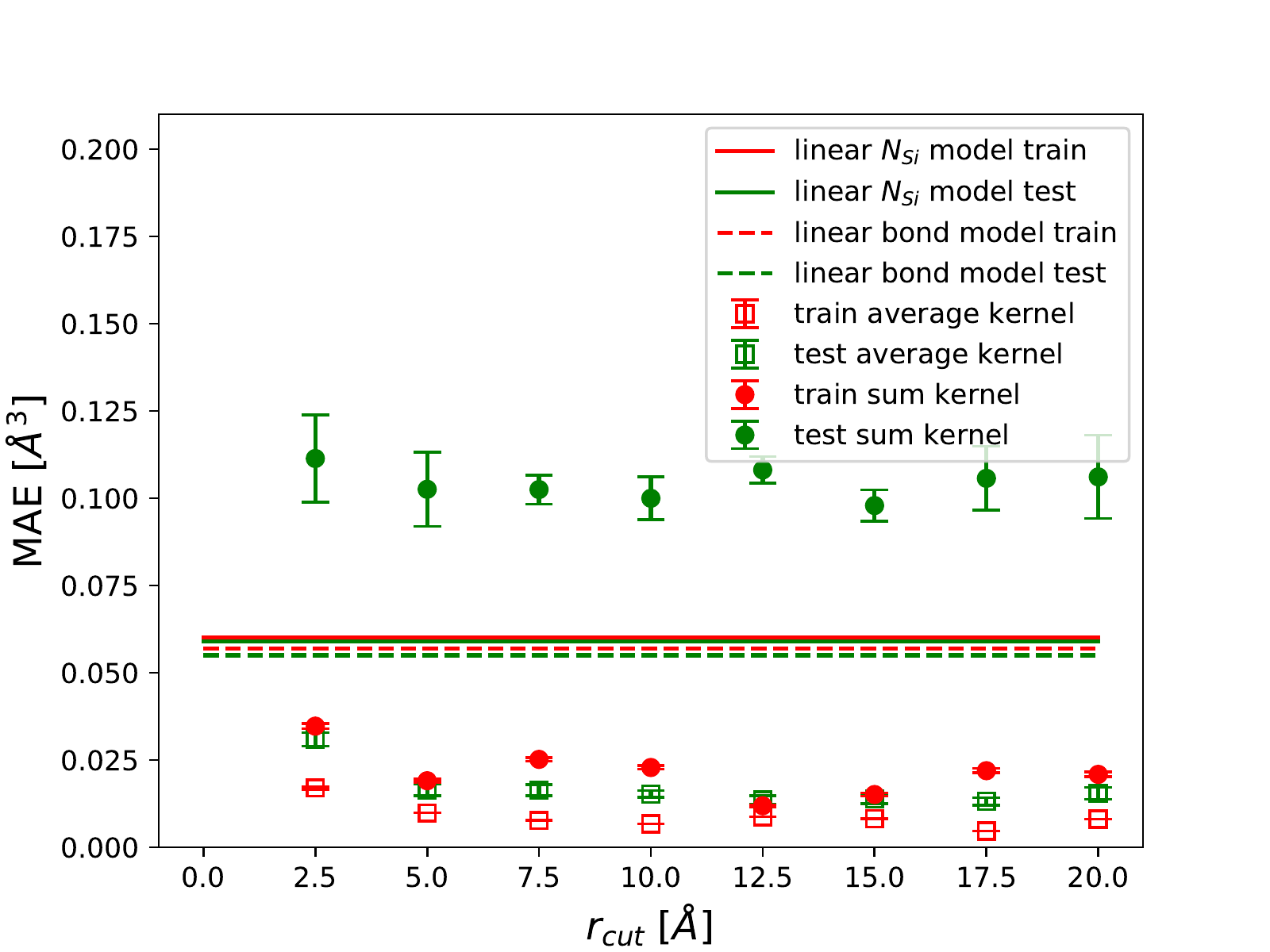}
    \end{subfigure}
    \begin{subfigure}{0.48\textwidth}
       \caption{}
    \includegraphics[width=0.99\textwidth]{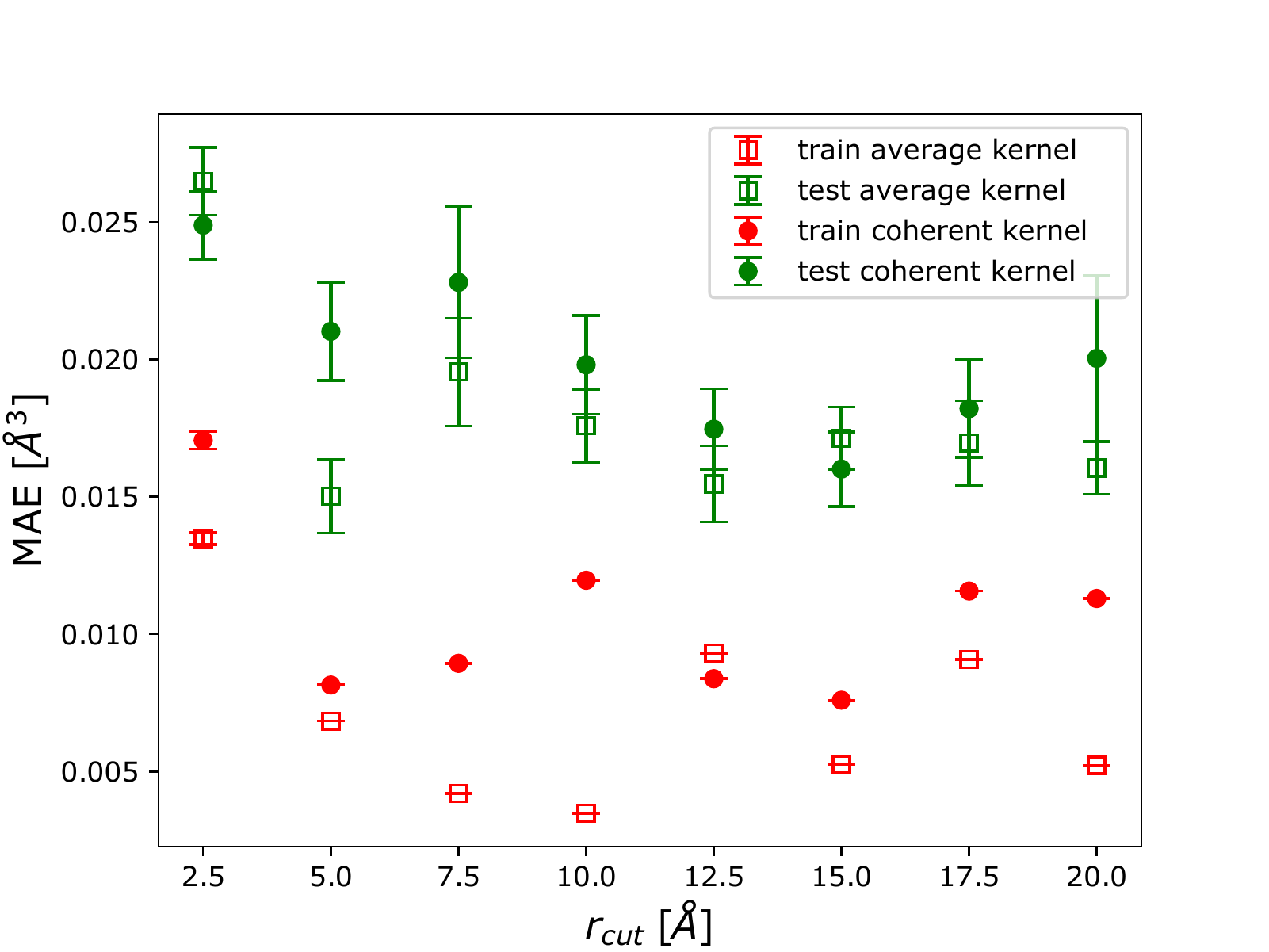}
    \end{subfigure}
     \caption{Average mean absolute error (MAE) of $\alpha/N_{Si}$ (the cluster polarizability divided by the number of silicon atoms) from various machine learning models and physical-based models versus the cut-off radius $r_{cut}$ that determines the size of the local chemical environment. Optimal regularization parameters were determined using five-fold cross-validation. Error bars indicate the standard error of the average MAE across the five training and validation sets used in the cross validation procedure.}
         \label{soapinterpolate}
\end{figure}

Up to this point, we only considered the ability of the SOAP approach to fit the calculated cluster polarizabilities. To investigate SOAP's capacity to predict polarizabilities of clusters that it was not trained on, we use k-fold cross validation~\cite{Rasmussen2005}. In this procedure, the clusters in the data set are randomly assigned to five sub-sets. Next, four sub-sets are used to train the ML approach and the fifth sub-set is used as the test set. This is done five times with each sub-set acting as test set once. We optimize the regularization parameter $\lambda$ to minimize the mean average error (MAE). The optimal parameters are listed in the appendix. The resulting MAE and its standard deviation as function of $r_\text{cut}$ are shown in Fig.~\ref{soapinterpolate} (a). The average kernel and the coherent kernel yield very similar results and are compared in Fig.~\ref{soapinterpolate} (b). Strikingly, the sum kernel model produces the largest MAE for the test set among all methods. In particular, the test set MAE is significantly larger than the training set MAE indicating poor capacity to predict polarizabilities. In contrast, the average kernel model yields the smallest test set MAE which is only slightly worse than the training set error. The coherent kernel model yields slightly worse predictions than the average kernel, with the biggest difference between the two occurring at $r_{cut}=7.5$~\AA. The MAE of the two physical-based models lies between those of the sum kernel and the average kernel. The different performances of the sum kernel and the average kernels originate from the different training procedures: the sum kernel model is trained on total cluster polarizabilities, while the average kernel is trained on the average polarizability per silicon atom, see Eq.~\eqref{squarekernel}. As a consequence, the sum kernel model is biased towards more accurate predictions for large clusters and is less accurate for small clusters. This can also be seen in Fig.~\ref{polfit}(c) which shows that the quality of the sum kernel fit improves for larger clusters. This has been observed before by Stocker et. al. \cite{Stocker2020} who argued that the intensive average kernel has the advantage of equally weighting small and large molecules, which is beneficial when learning quantities over a large range of cluster sizes. Interestingly, the average kernel performs somewhat better than the coherent kernel suggesting that a model of the cluster polarizability that can be expressed as a sum of atomic contributions constitutes a better representation of the system's dielectric response.

\begin{figure}
    \begin{subfigure}[b]{0.48\textwidth}
    \caption{}
     \includegraphics[width=1\linewidth]{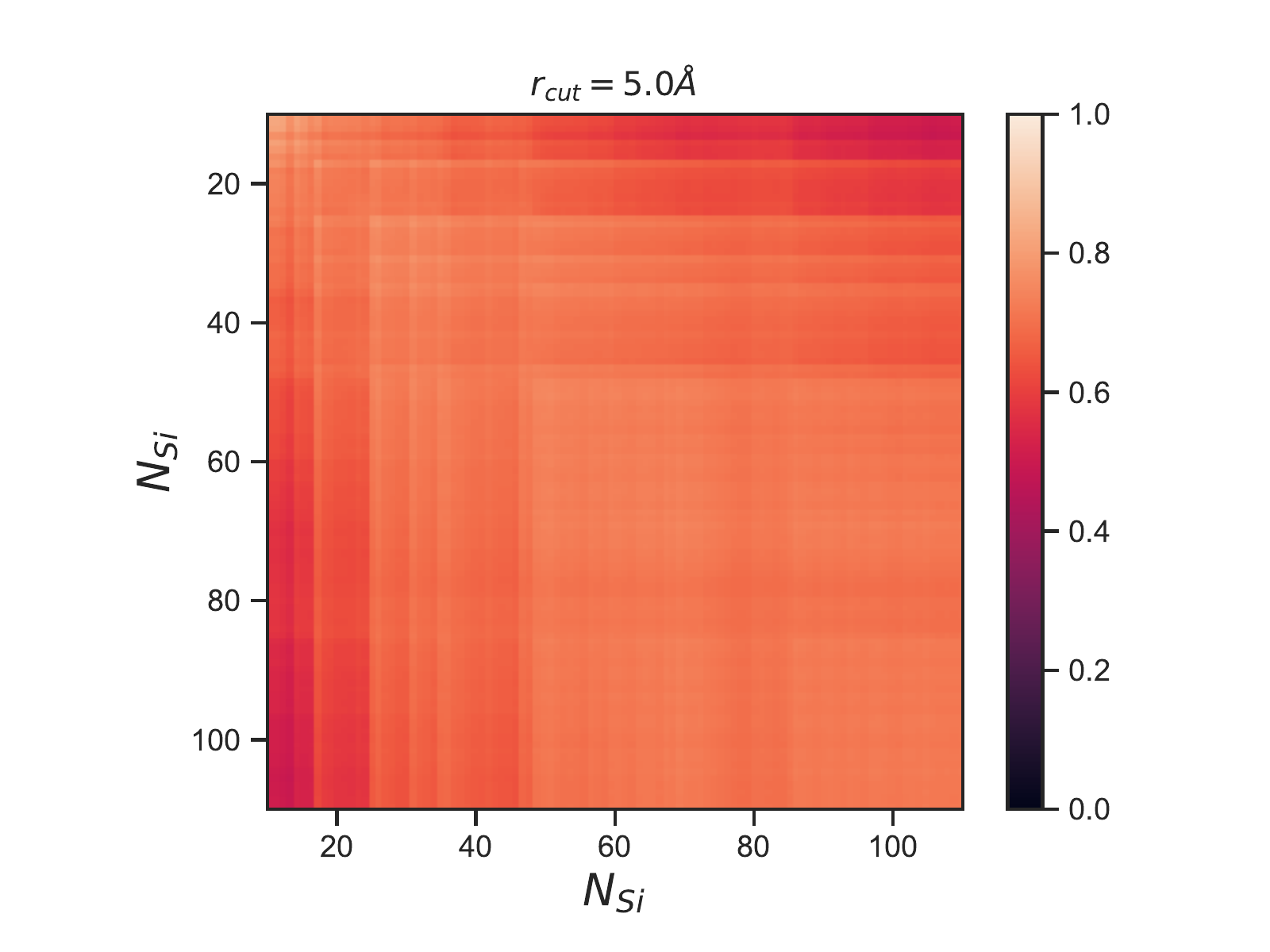}
    \end{subfigure}
    \begin{subfigure}[b]{0.48\textwidth}
    \caption{}
    \includegraphics[width=1\linewidth]{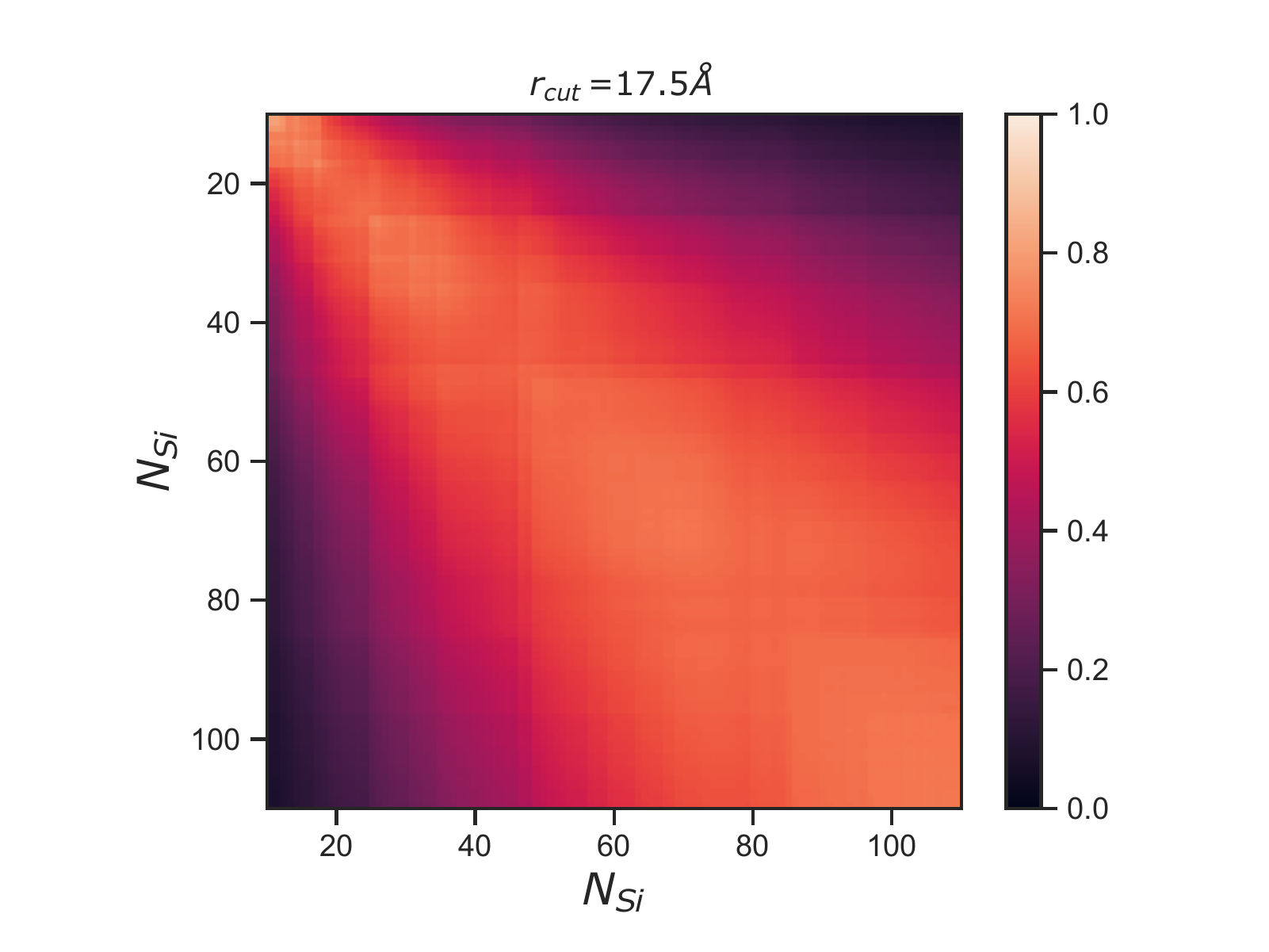}
    \end{subfigure}
    \caption{Matrix elements of the average SOAP kernel $K^{av}_{IJ}$, see Eq.~\eqref{eq:Kav}, for cut-offs $r_{cut}$ of 5.0 \AA\; (a) and 17.5 \AA\; (b). Note the rapid decay of the matrix elements along rows and columns when a large cut-off radius is used.}
    \label{kernelmatrices}
\end{figure}

Fig.~\ref{soapinterpolate} also shows that the minimum test set MAE for the average kernel and the coherent kernel is obtained for $r_{cut}=17.5$~$\text{\AA}$, while for the sum kernel the minimum is achieved for $r_{cut}=15.0$~$\text{\AA}$. Interestingly, neither kernel benefits significantly from increasing $r_{cut}$ beyond 5~\AA. To understand this finding, we compare the elements of the average kernel matrix for $r_{cut}=5.0$ \AA\ and $r_{cut}=17.5$ \AA, see Fig.~\ref{kernelmatrices}. For the smaller cutoff, the kernel matrix decays slowly along the rows and columns of the kernel matrix. In contrast, the decay is significantly more pronounced for the larger cutoff suggesting that a smaller cutoff facilitates the recognition of similar chemical environments in clusters of different size. This is not surprising because for large cutoffs the chemical environment contains a significant amount of vacuum for small clusters, but not for large clusters. 

\begin{figure}
    \centering
    \includegraphics[scale=0.48]{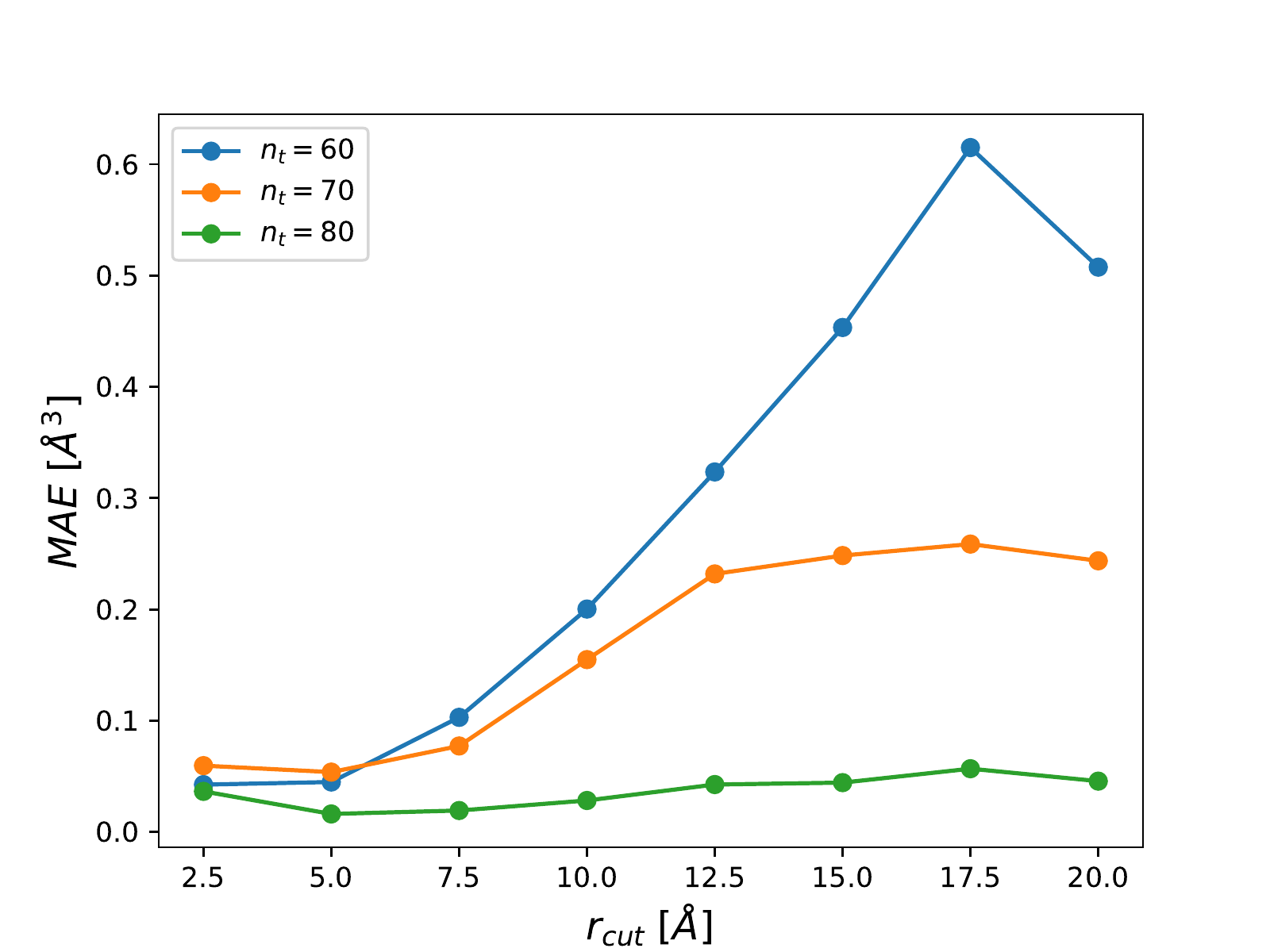}
    \caption{Average kernel test set error as a function of SOAP cut-off. The smallest $n_t$ clusters were included in the training set for each curve. The test set consists of the remaining $100-n_t$ clusters.}
    \label{extrap}
\end{figure}

Next, we explore the ability of the ML approach to predict polarizabilities of large clusters based on a training set of small clusters. For this, we train the average kernel on the 60, 70 or 80 smallest clusters and then predict the polarizabilities of the remaining large clusters in the data set. Fig.~\ref{extrap} shows the resulting test set MAE as function of the cutoff radius. All curves exhibit a minimum at small cut-offs near $r_{cut}=5$~\AA\ and the smallest MAE is obtained for the largest training set. For the smaller training sets ($n_t=60$ or 70) the MAE increases rapidly as the cutoff is increased, while for the largest training set the increase is mild (and another minimum is found at $r_{cut}=17.5$~\AA). Similar to our findings in the k-fold cross validation, this shows that it is not beneficial to increase the cut-off radius beyond a certain value. 

Figure~6 (a)-(c) compare the predictions of the average kernel with $r_\text{cut}=5$~\AA\; with the calculated RPA polarizabilities per silicon atom. For all three training set sizes, the ML model captures the qualitative trends. For $n_t=60$, the average kernel correctly predicts the increase of $\alpha/N_{Si}$ at $N_{Si}=70$ and also the decrease starting at $N_{Si}=80$. While the ML models underestimate the polarizabilities per Si atom for large clusters when $n_t=60$ and $n_t=70$, good quantitative agreement is achieved for $n_t=80$.

\begin{figure}
    \centering
    \captionsetup[subfigure]{labelformat=empty}
    \begin{subfigure}[b]{0.0\textwidth}
   \caption{\label{nextr1}}
    \end{subfigure}
    \captionsetup[subfigure]{labelformat=empty}
    \begin{subfigure}[b]{0.0\textwidth}
    \caption{\label{nextr2}}
    \end{subfigure}
    \captionsetup[subfigure]{labelformat=empty}
     \begin{subfigure}[b]{0.48\textwidth}
    \includegraphics[width=1.0\textwidth]{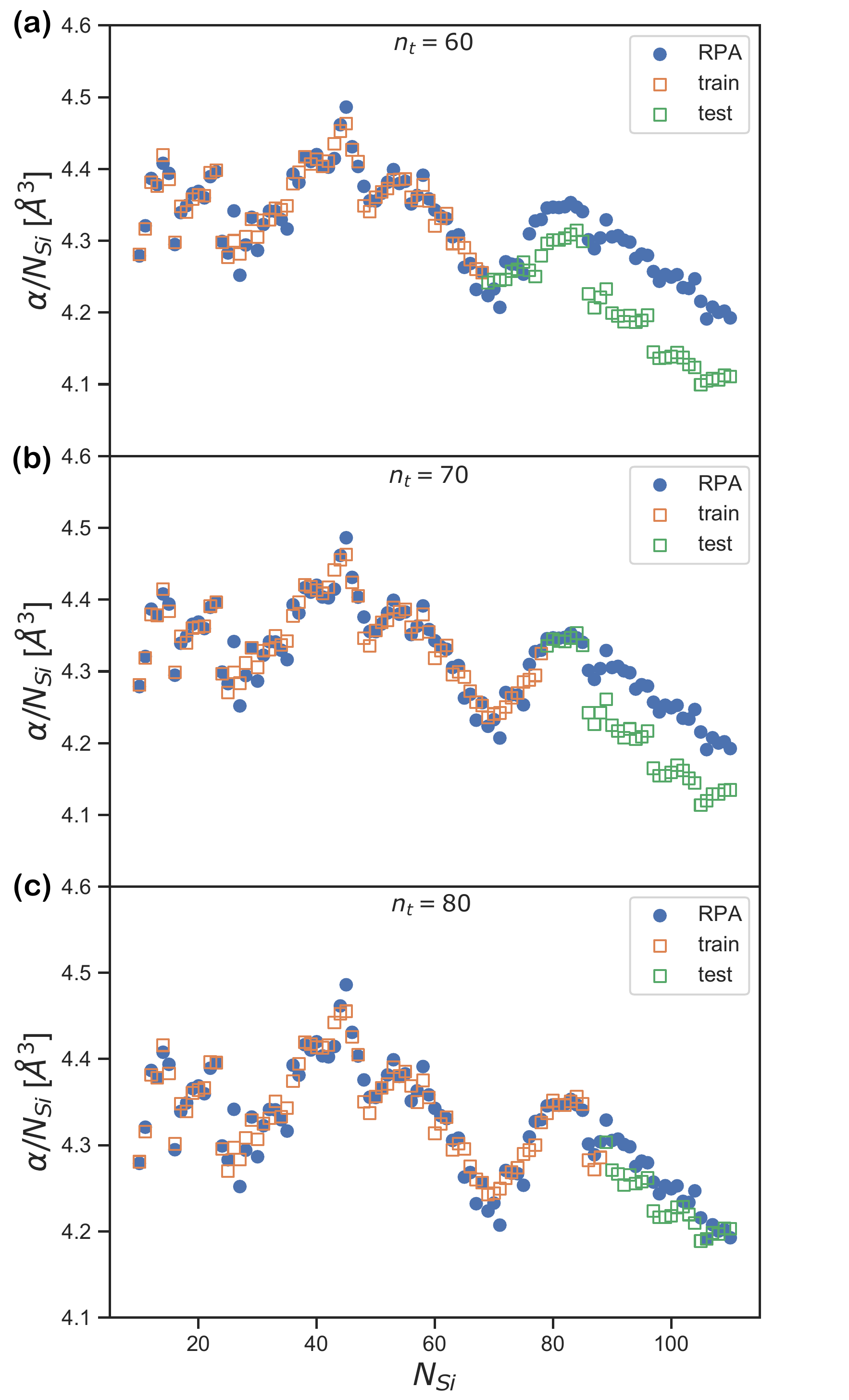}
    \caption{\label{nextr3}}
    \end{subfigure}
       \caption{Comparison of RPA results for $\alpha/N_{Si}$ (cluster polarizability divided by the number of silicon atoms) and training and test set predictions of the average kernel model. The training set consists of the (a) $n_t=60$, (b) $n_t=70$ and (c) $n_t=80$ smallest clusters and the test set contains the remaning $100-n_t$ large clusters.}
    \label{nextr}
\end{figure}

\begin{figure}
    \centering
    \includegraphics[scale=0.48]{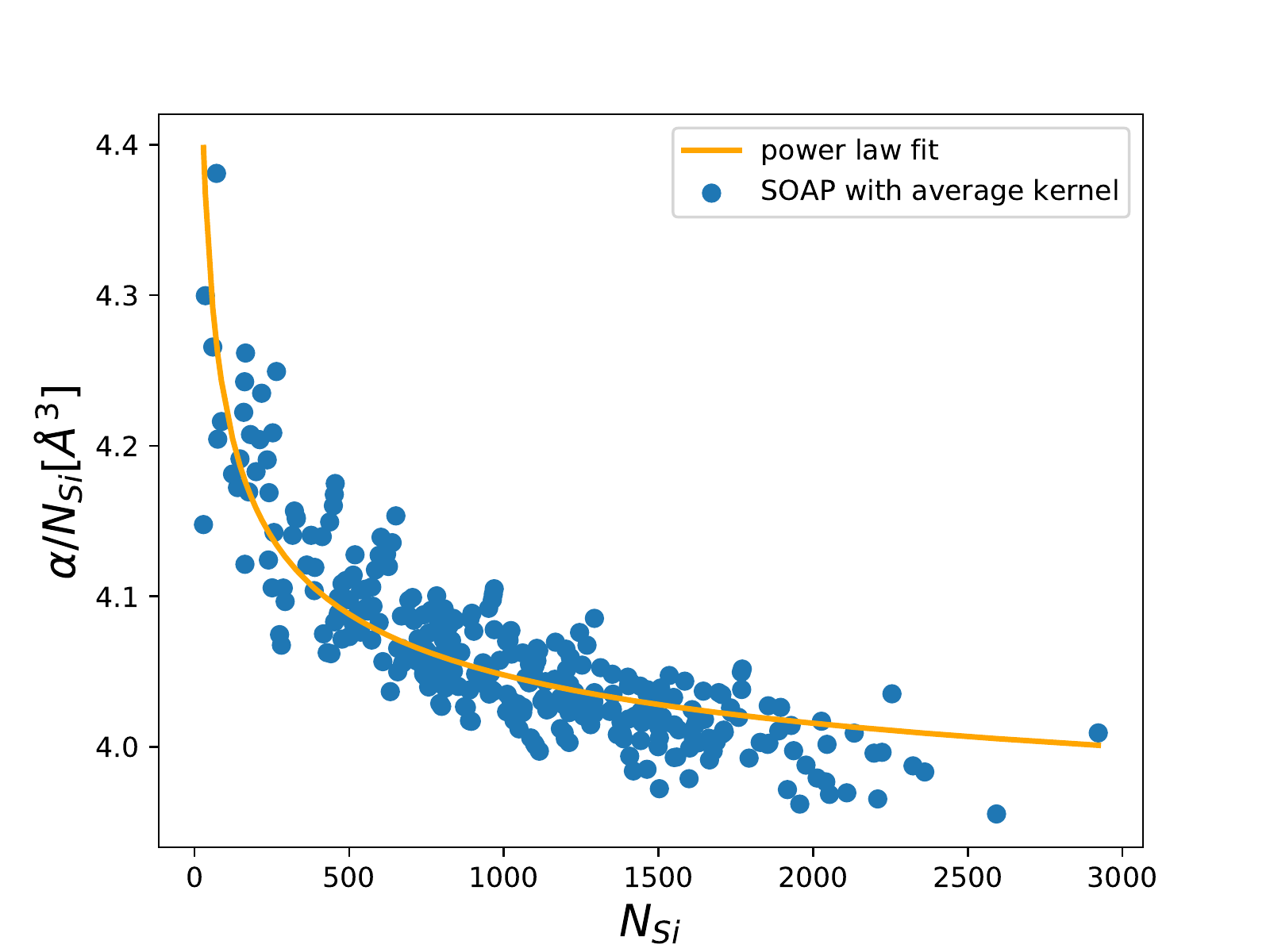}
    \caption{Cluster polarizability divided by the number of silicon atoms for all clusters of Silicon Quantum Dot database~\cite{barnard2015silicon} from the average SOAP kernel model with $r_{cut}=5.0$\;\AA. }
    \label{siquantumdot}
\end{figure}

Finally, we train the average kernel model on the entire data set (using $r_{cut}=5$~\AA) and predict the average polarizabilities of the entire Silicon Quantum Dot data set containing clusters with up to 3000 silicon atoms~\cite{barnard2015silicon}. The results are shown in Fig.~\ref{siquantumdot}. It can be observed that the polarizability per Si atom converges slowly to its bulk limit as $N_{Si}$ increases and there is significant scatter in the results. The scatter in $\alpha/N_{Si}$ reflects the different $N_H/N_{Si}$ ratios and different environments present in the clusters. To understand the slow convergence to the bulk value, note that the number of silicon atoms scales with the cluster volume, while the number of hydrogen atoms is roughly proportional to the surface area. This suggests that $\alpha/N_{Si}$ should
be proportional to the inverse radius of the cluster or, equivalently, to  $1/N_{Si}^{1/3}$. Indeed, Fig.~\ref{siquantumdot} shows that the ML predictions are well described by the function $a+b/N_{Si}^{1/3}$ with $a=3.89$~\AA$^3$ and $b=1.55$ obtained from a least-squares fit. The value of $a$ agrees well with the RPA atomic polarizability of bulk silicon of  3.77~\AA$^3$\cite{Hybertsen1987}.

\begin{figure}[H]
    \centering
    \begin{subfigure}{0.48\textwidth}
        \caption{}
        \includegraphics[width=0.99\textwidth]{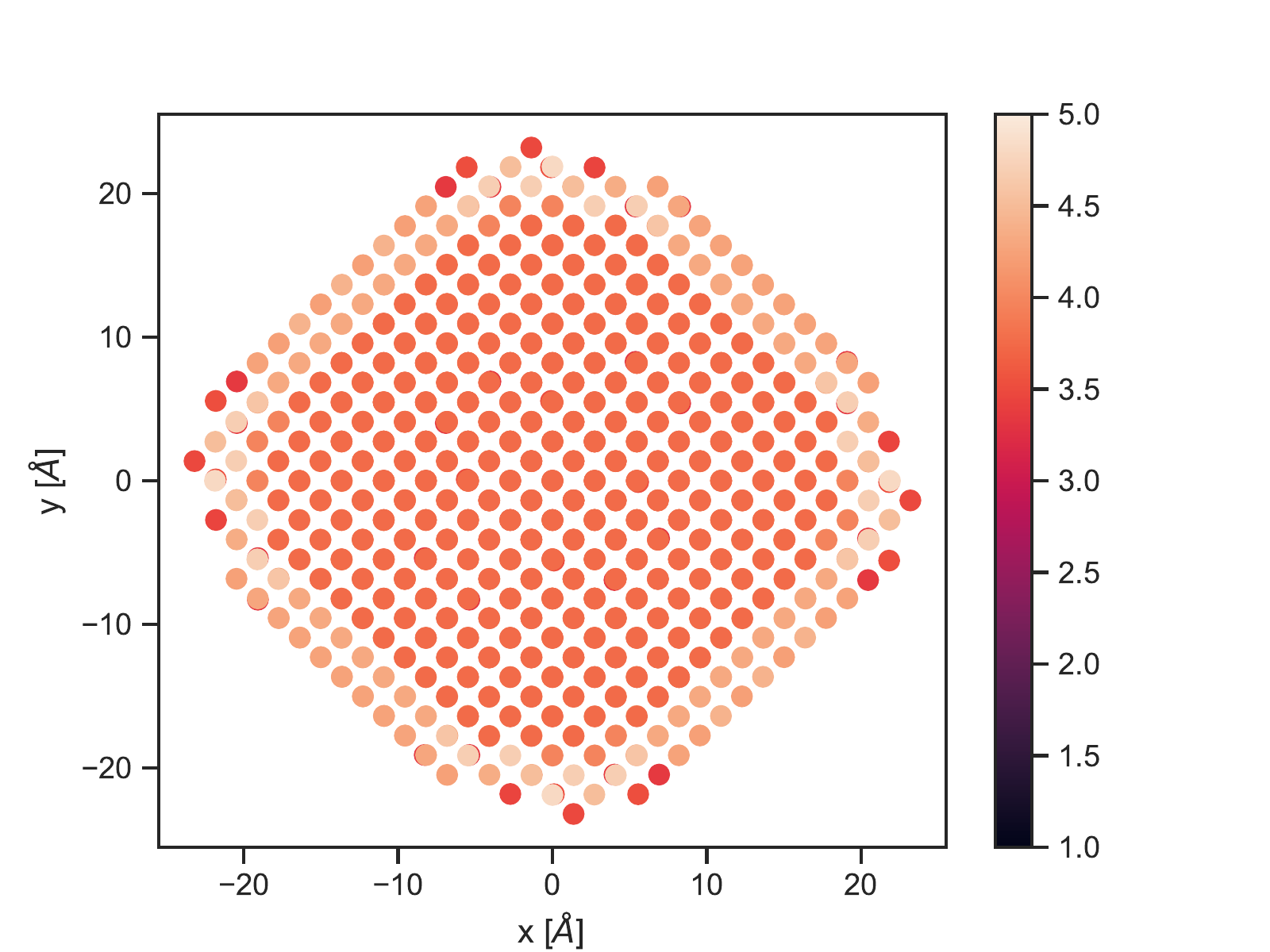}
    \end{subfigure}
      \begin{subfigure}{0.48\textwidth}
        \caption{}
        \includegraphics[width=0.99\textwidth]{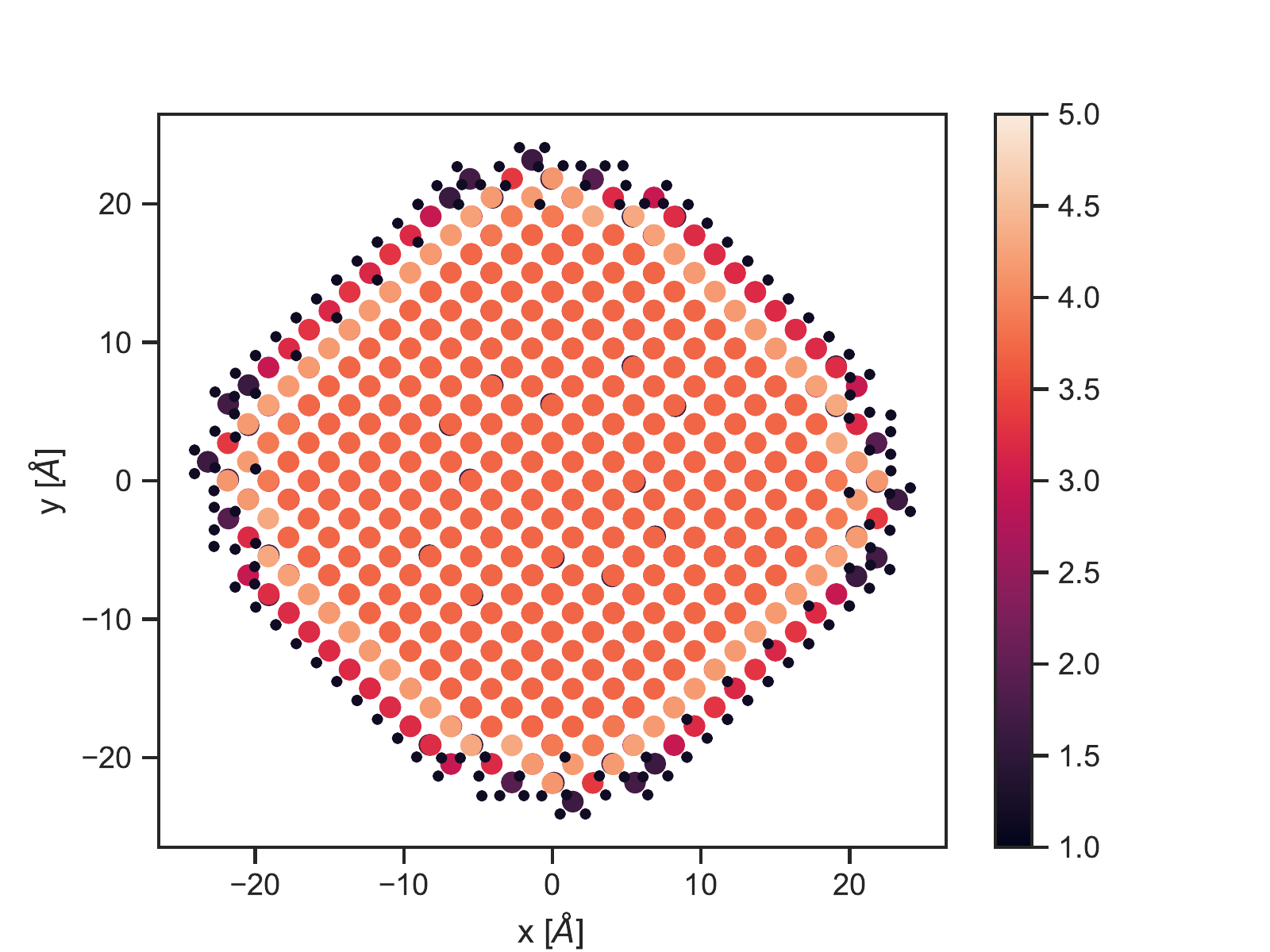}
    \end{subfigure}
    \caption{Atomic polarizabilities of the Si$_{2109}$H$_{604}$ cluster obtained from the SOAP average kernel method. Shown is a cross section through the center of the cluster. (a) Atomic polarizabilities when only silicon chemical environments are used. (b) Atomic polarizabilities when both silicon and hydrogen chemical environments are used. For hydrogen environments $r_{cut}=1.6$~\AA\; was used and for silicon environments $r_{cut}=5.0$~\AA\; was used. Large dots represent silicon atoms and small dots represent hydrogen atoms.}
    \label{crosssection}
\end{figure}

Additional insights can be obtained by analyzing the atomic polarizabilities obtained from the SOAP average kernel method, see Eq.~\eqref{eq:atomic}. Fig.~\ref{crosssection} shows the atomic polarizabilities of a Si$_{2109}$H$_{604}$ cluster. In Fig.~\ref{crosssection} (a) only local chemical environments of silicon atoms are considered (and the effect of the hydrogen atoms is captured indirectly through their influence on the silicon chemical environments). Silicon atoms in the center of the cluster have a polarizability of  3.76~\AA$^3$, in excellent agreement with value extracted from bulk calculations of 3.77~\AA$^3$~\cite{Hybertsen1987}. The polarizability of the silicon atoms in the two surface layers is larger, sometimes as large as 5~\AA$^3$. The reason for this increase is that the surface silicon atoms are bonded to hydrogen atoms and their atomic polarizability is effectively the sum of the silicon and hydrogen contributions. To disentangle contributions from silicon and hydrogen atoms to the cluster polarizability, Fig.~\ref{crosssection} shows the atomic polarizabilities from a calculation that explicitly takes chemical environments of hydrogen atoms into account. Interestingly, the results suggest that the atomic polarizability of subsurface silicon atoms is larger than the bulk value, but the polarizability of surface silicon atoms (which are bonded to hydrogens) is smaller. The average atomic polarizability of the silicon atoms is found to be 3.63~\AA$^3$. This is in agreement with the results of Mochizuki et. al~\cite{MOCHIZUKI2001451}, who predicted that the bulk limit of the silicon atomic polarizability is approached from below. 

\section{Conclusions}
In this work, we have demonstrated that machine learning models based on the smooth overlap of atomic positions (SOAP) descriptor can be used to accurately and efficiently predict polarizabilities of large hydrogenated silicon clusters. Using the random phase approximation, we calculated the polarizabilities of a set of hydrogenated silicon clusters containing between 10 and 110 silicon atoms. We then assessed the ability of three machine learning models (one using the sum kernel, one using the average kernel and one the coherent kernel) to fit the calculated polarizabilities and find that all three models perform well when the local environment includes nearest neighbour atoms only. Increasing the size of the environment improves the quality of the fit. Next, we investigated the ability of the machine learning models to predict polarizabilities of clusters that are not in the training set. Using k-fold cross validation, we find that the average kernel performs significantly better than the sum kernel and that the predictions only weakly depend on the size of the chemical environment. We also tested the predictive power of the average kernel when it is trained on small clusters only and find that quantitative accuracy can be achieved if the training set is sufficiently large. Finally, we use the average kernel approach to predict the polarizabilities of hydrogenated silicon atoms with up to 3000 silicon atoms and find that the results approach the correct bulk limit. The ability to efficiently calculate polarizabilities of large clusters paves the way towards using machine learning for excited-state properties of these systems. For example, the static density-density response function (from which the polarizability is calculated) is a key ingredient for calculating quasiparticle properties within the GW approach (typically when used in conjunction with a generalized plasmon-pole model) and also for calculating optical properties by solving the Bethe-Salpeter equation. Symmetry-adapted kernel regression could be used to straightforwardly generalise our models to predict the full polarisability tensor~\cite{Grisafi2018}.

\section{Data availability statement}
The data that support the findings of this study are available upon reasonable request from the authors.

\section{Acknowledgements}
This work was supported through a studentship in the Centre for Doctoral Training on Theory and Simulation of Materials at Imperial College London funded by the EPSRC (EP/L015579/1). We acknowledge the Thomas Young Centre under grant number TYC-101. This work used the ARCHER UK National Supercomputing Service (http://www.archer.ac.uk), and the Imperial College London High-Performance Computing Facility.
\vfill\clearpage

\bibliography{bibliography.bib}

\section{Appendix}
\begin{table}[h]
    \centering
    \caption{Regularization paramaters $\lambda_{av}$ and $\lambda_{sum}$ determined from k-fold cross validation at different cut-off radii $r_{cut}$. }
    \begin{tabular}{|c|c|c|c|}
    \hline
    $r_{cut}$ [\AA]&  $\lambda_{av}$ & $\lambda_{sum}$ & $\lambda_{coh} $ \\
    \hline
       $2.5$&   $10^{-8}$ &$10^{-8}$ & $10^{-8}$\\
       $5.0$ & $10^{-5}$ & $0.0001$&$10^{-6}$ \\
         $7.5$ & $10^{-5}$ &$0.01$& $10^{-5}$ \\
      $10.0$ &  $0.0001$ &$0.01$& $0.0001$ \\
       $12.5$ & $10^{-5}$ & $10^{-6}$ & $10^{-5}$ \\
       $15.0$ & $10^{-5}$ &$10^{-5}$& $10^{-5}$ \\
       $17.5$ & $0.0001$ &  $0.001$ & $0.0001$ \\
       $20.0$ &$0.0001$ &$0.001$ & $0.0001$ \\
\hline
    \end{tabular}

    \label{tab:my_label}
    
\end{table}
\end{document}